\documentclass[floatfix,showpacs,prd,aps,tightenlines,superscriptaddress]{revtex4-1}
\usepackage{amsmath}
\usepackage{graphicx}
\usepackage{psfrag}
\usepackage{color}
\usepackage{dcolumn}
\usepackage{bm}
\usepackage{longtable}

\def\edth{{\rlap{$\partial$}\raise0.3em\hbox{$-$}}}

\newcommand{\bea}{\begin{eqnarray}}
\newcommand{\eea}{\end{eqnarray}}

\usepackage[mathscr]{eucal}
\usepackage{mathrsfs}

\newcommand{\scri}{\mathscr{I}}
\begin{document}

\title{Perturbative extraction of gravitational waveforms generated
with Numerical Relativity}

\author{Hiroyuki Nakano}
\affiliation{Department of Physics, Kyoto University, Kyoto 606-8502, Japan.}
\affiliation{Center for Computational Relativity and Gravitation,\\
and School of Mathematical Sciences, Rochester Institute of
Technology, 85 Lomb Memorial Drive, Rochester, New York 14623}

\author{James Healy}
\affiliation{Center for Computational Relativity and Gravitation,\\
and School of Mathematical Sciences, Rochester Institute of
Technology, 85 Lomb Memorial Drive, Rochester, New York 14623}

\author{Carlos O. Lousto}
\affiliation{Center for Computational Relativity and Gravitation,\\
and School of Mathematical Sciences, Rochester Institute of
Technology, 85 Lomb Memorial Drive, Rochester, New York 14623}

\author{Yosef Zlochower}
\affiliation{Center for Computational Relativity and Gravitation,\\
and School of Mathematical Sciences, Rochester Institute of
Technology, 85 Lomb Memorial Drive, Rochester, New York 14623}

\begin{abstract}
We derive an analytical expression for extracting the
gravitational waveforms at null infinity using the Weyl scalar $\psi_4$
measured at a finite radius. Our expression is based on a series
solution in orders of 1/r to the equations for gravitational perturbations
 about a spinning black hole.  We compute this expression to order $1/r^2$ and include
the spin parameter $a$ of the Kerr background.  We test the accuracy
of this extraction procedure by measuring the waveform for a merging
black-hole binary at ten different extraction radii (in the range
$r/M=75\--190)$ and for three different resolutions in the convergence
regime.  We find that the extraction formula provides a set of values
for the radiated energy and momenta that at finite extraction radii
converges towards the expected values with increasing resolution,
which is not the case for the `raw' waveform at finite radius. 
We also examine the
phase and amplitude errors in the waveform as a function of observer
location and again observe the benefits of using our extraction
formula.  The leading corrections to the phase are ${\cal
O}(1/r)$ and to the amplitude are ${\cal O}(1/r^2)$. This method
provides a simple and practical way of estimating the waveform at
infinity, and may be especially useful for scenarios such as well
separated binaries, where the radiation zone is far from the sources,
that would otherwise require extended simulation grids in order
 to extrapolate
the `raw' waveform to infinity.  Thus this method saves important
computational resources and provides an estimate of errors.
\end{abstract}

 \maketitle

\section{Introduction}

Perturbation theory about black-hole backgrounds and fully nonlinear
numerical simulations of the Einstein field equations provide
complementary approaches to solving important problems in Relativity.
A few examples of the synergy created by using the two together
include the use of perturbative boundary techniques for a fully
nonlinear simulation far from the sources as a way of propagating most
of the radiation out the simulation domain \cite{Abrahams97a} and the
now classical Lazarus approach \cite{Baker:2000zh, Baker:2001nu,
Baker:2001sf, Baker:2002qf, Baker:2003ds, Campanelli:2005ia} that
extracted spatial information from a short lived full numerical
evolution to provide initial data for a subsequent perturbative
evolution of a single rotating black hole.

With the breakthroughs in numerical
relativity~\cite{Pretorius:2005gq,Campanelli:2005dd, Baker:2005vv},
complete simulations of inspiraling black-hole binaries became
possible. However, even in this case, the spacetime far from the
sources (more precisely, in the wavezone) can be described by black
hole perturbation theory.
Here we will exploit this fact to {\it analytically} propagate the
waveform from a fully nonlinear simulation, but measured at finite
distance from the sources, to null infinity.

Over the last few years several
of these waveform extraction techniques have been developed.
The most straightforward strategy would be to have the numerical
domain extend very far from the sources and extrapolate the waveform
measured at far distances to infinity.
 This can be achieved at reasonable computational efficiency using 
pseudo-spectral decomposition of
the fields \cite{Scheel:2006gg,Scheel:2008rj},
or by using multi-patch techniques~\cite{Pazos:2009vb, Pollney:2009yz}.

A more sophisticated waveform extraction technique, and one that
produces a true gauge invariant signal, is Cauchy-Characteristic
extraction (CCE)~\cite{Reisswig:2009rx, Babiuc:2010ze, Taylor:2013zia}. In this technique, the metric and its derivatives on a
timelike worldtube are used as inner boundary data for a subsequent
characteristic evolution. As the characteristic evolution includes
null infinity, the waveform obtained is exact (up to truncation
error). A complementary approach to CCE
is to evolve the spacetime on surfaces that are spacelike in the
interior but asymptote to null slices that intersect, $\scri^+$, null
infinity~\cite{Vano-Vinuales:2014ada, Vano-Vinuales:2014koa}.

An alternative extrapolation method consists of using the results of
perturbation theory to  propagate waveforms obtained at finite radii
(but in the radiation zone) to infinity. Treating the background
spacetime as a perturbation of Schwarzschild, which will be accurate
in the wavezone, leads to a simple explicit formula  relating $\psi_4$
at infinity with the finite radius $\psi_4$ and its time integral. For
more details, see Ref.~\cite{Lousto:2010qx}, Eq.~(53). This method has
been proven to correct for the next-to-leading  $1/R_{\rm obs}$ term in
$R_{\rm obs} \psi_4(t,R_{\rm obs})$~\cite{Babiuc:2010ze,Lousto:2012su} using
only a single observer radius and displays a significantly reduced
level of extrapolation noise, when compared to the standard polynomial
extrapolation.  The errors produced by this method can be estimated by
applying it to different extraction radii. We applied this method to the
$q=10$ case in Ref.~\cite{Hinder:2013oqa}
and found good agreement (but with significantly reduced 
noise) between the perturbative and standard extrapolation technique 
used in this paper.

In this paper we expand upon this method by including higher-order
[${\cal O}(1/r^2)$] and rotation effects for extracting or extrapolating
fields from an intermediate distance to infinity via a perturbative,
analytic expansion. This method is relatively simple to implement, yet
it is accurate enough for most applications. This includes cases of
large-separation  binaries with long orbital periods leading to wave
zones extending beyond several thousand $M$ (see for instance
Ref.~\cite{Lousto:2013oza} where evolution of a binary separated by
$100M$ led to waveforms with $6400M$ periods), as well as studies of
the scattering of two black holes starting far apart to measure small
scattering angles \cite{Damour:2014afa} and high energy collision of
black holes \cite{Sperhake:2008ga}, which require extractions at large
distances from the sources.  Another circumstance when this extraction
method can be of use is when more physical scales need to be resolved.
Such is the case when matter surrounds black holes
\cite{Zilhao:2014ida} or when one tries to simulate a hybrid systems
involving neutron stars and black holes or binary neutron stars
\cite{Kyutoku:2014yba}.

The paper is organized as follows. Section~\ref{sec:Pert}
discusses the extraction of gravitational waves
propagating as a perturbation on the asymptotic Schwarzschild background
with ${\cal O}(1/r)$ corrections included in Sub-Sec.~\ref{sec:1st} 
and ${\cal O}(1/r^2)$ corrections in Sub-Sec.~\ref{sec:2nd}.
In Sec.~\ref{sec:PertSpin} we include the effects of the spin in the
background. Linear corrections in the spin in Sub-Sec.~\ref{sec:Kerr1}. 
In Sub-Sec.~\ref{sec:Lazarus} we correct the extractions of the
Weyl scalar $\psi_4$ for a nonconventional choice of the tetrad used
in full numerical simulations. While in Sub-Sec.~\ref{sec:Improved}
we collect together a definitive formula to include all effects
together. This formula is capable of extracting numerical $\psi_4$
relatively close to the sources and extrapolate waveforms to infinity
with accuracy, particularly for its phase and amplitude. 
Section~\ref{sec:Lazarus_E} contains explicit
expressions for the radiated energy and momenta (along the z-axis)
based on the extrapolated waveforms. In Sec.~\ref{sec:FN} we apply
those equations into a case study of full numerical evolution of
binary black holes. We choose ten extraction radii in the intermediate
radiation region and evolve with three different resolutions in the
convergence regime to study the effects of finite resolution on the
extrapolated quantities. We finish the paper with a brief discussion
in Sec.~\ref{sec:discussion} of the range of applicability of our results.

\section{Perturbation in a nonspinning background}\label{sec:Pert}

In this section we derive expressions relating the
Regge-Wheeler-Zerilli (RWZ)~\cite{Regge:1957td,Zerilli:1971wd}
 functions at finite radius to their values on
$\scri^+$ in the Schwarzschild (mass $M$) black hole perturbation.
Then, using these expressions, we derive expressions for
the Weyl scalar $\psi_4$ at $\scri^+$ based on its values at finite
radii. We always work in the first order perturbative regime, i.e.
no quadratic terms in the perturbations around the black hole
background are included, and expand the solutions asymptotically in powers
of $1/r$.

\subsection{First-order corrections: $(1/r)$-terms}\label{sec:1st}

The Weyl scalar, $\psi_4$, in an asymptotically flat tetrad, like
Kinnersley's~\cite{Kinnersley_1969}, is related to the strain
at large radii  by
\bea
\lim_{r\to\infty} r\psi_4 &=&\lim_{r\to\infty} r( \ddot{h}_{+} -
i\,\ddot{h}_{\times} )
\,.
\label{eq:psi4_hp_hc}
\eea
Similarly,
the RWZ even and odd parity functions are related to the strain on
at large radii by
\bea
h_{+} - i\,h_{\times} =
\sum_{\ell m} \frac{\sqrt{(\ell-1)\ell(\ell+1)(\ell+2)}}{2r} 
\left(
\Psi_{\ell m}^{\rm (even)} - i\, \Psi_{\ell m}^{\rm (odd)}
\right)
{}_{-2}Y_{\ell m}
\,,
\label{eq:RWZwaves}
\eea
where $\Psi_{\ell m}^{\rm (even)}$ and $\Psi_{\ell m}^{\rm (odd)}$
are the even and odd parity wave functions, respectively,
and ${}_{-2}Y_{\ell m}$ denotes the spin($-2$)-weighted spherical harmonics
(see, e.g., review papers~\cite{Nakamura87,Lousto:2005ip,Lousto:2005xu,Nagar:2005ea}).

The asymptotic values of $\psi_4$ and the RWZ wavefunctions
can be related to their values at finite radii by examining the
asymptotic behavior of the relevant wave equations.
For the RWZ wave equations we get,
\bea
\Psi_{\ell m}^{\rm (even/odd)} &=& H_{\ell m}^{\rm (even/odd)}(t-r^*) 
+ \frac{\ell(\ell+1)}{2\,r} \int dt H_{\ell m}^{\rm (even/odd)}(t-r^*) 
+ {\cal O}(1/r^2) \,,
\label{eq:asymtRWZ}
\eea
for general $\ell$ modes, where $H_{\ell m}$ is the strain observed
at infinity,
and $r^*=r+2M\ln[r/(2M)-1]$. An error due to finite extraction radii
arises from the integral term in Eq.~(\ref{eq:asymtRWZ}).
Inverting the above relation, we have~\cite{Lousto:2010qx}
\bea
\left. \Psi_{\ell m}^{\rm (even/odd)}\right|_{r=\infty} &=& \Psi_{\ell m}^{\rm (even/odd)}(t,r)
- \frac{\ell(\ell+1)}{2\,r} \int dt \Psi_{\ell m}^{\rm (even/odd)}(t,r)
+ {\cal O}(1/r^2) \,. 
\eea
Similarly,
if the Weyl scalar,
\bea
\psi_4 = \sum_{\ell m} \psi_4^{\ell m} {}_{-2}Y_{\ell m} \,,
\eea
satisfies the Teukolsky equation~\cite{Teukolsky:1973ha}
in the Schwarzschild background spacetime, then the asymptotic
behavior of $\psi_4^{\ell m}$ is
given by
\bea
r\,\psi_4^{\ell m} &=& \ddot {\tilde H}_{\ell m} (t-r^*) 
+ \frac{(\ell -1)(\ell +2)}{2\,r} \dot {\tilde H}_{\ell m} (t-r^*)
+ {\cal O}(1/r^2) \,,
\label{eq:asymtPsi4}
\eea
where the difference between ${\tilde H}_{\ell m}$ 
and $H_{\ell m}=H_{\ell m}^{\rm (even)} - iH_{\ell m}^{\rm (odd)}$
defined from Eq.~(\ref{eq:asymtRWZ}) is only a numerical factor,
and we have the relation by using Eqs.~(\ref{eq:psi4_hp_hc}) and~(\ref{eq:RWZwaves}) as
\bea
{\tilde H}_{\ell m} = \frac{\sqrt{(\ell-1)\ell(\ell+1)(\ell+2)}}{2} 
H_{\ell m} \,.
\eea
Inverting Eq.~(\ref{eq:asymtPsi4}), we get
\bea\label{eq:Schw1}
\left. r\,\psi_4^{\ell m}\right|_{r=\infty} &=& r\,\psi_4^{\ell m}(t,r)
- \frac{(\ell -1)(\ell +2)}{2\,r} \int dt \,[r\,\psi_4^{\ell m}(t,r)]
+ {\cal O}(1/r^2) \,.
\eea

To see the phase and amplitude corrections by using the above formula,
we assume 
\bea
H_{\ell m}^{\rm (even/odd)} (t-r^*) = A_{\ell m} \exp(-i \omega_{\ell m} (t-r^*)) \,,
\eea
in Eq.~(\ref{eq:asymtRWZ}).
Then, the RWZ functions at a finite extraction radius are given by~\cite{Nakano:2015rda}
\bea
\Psi_{\ell m}^{\rm (even/odd)} &=& A_{\ell m}
\left[1 + \frac{1}{2}\left(\frac{\ell (\ell+1)}{2 \omega_{\ell m} r}\right)^2 + {\cal O}(1/r^4) \right] \exp(-i \omega_{\ell m} (t-r^*))
\exp(i\,\delta \phi_{\ell m}) + {\cal O}(1/r^2) \,,
\label{eq:APcorrection}
\eea
where $\delta \phi_{\ell m}$ is defined as
\bea
\sin \delta \phi_{\ell m}
&=& \frac{\ell (\ell+1)}{2 \omega_{\ell m} r} + {\cal O}(1/r^2) \,.
\label{eq:PHcorrection}
\eea
Therefore, the phase correction from the perturbative formula has ${\cal O}(1/r)$.
On the other hand, from Eq.~(\ref{eq:APcorrection}) the amplitude correction will be 
${\cal O}(1/r^2)$ which we have ignored here.
This result is consistent with Refs.~\cite{Hannam:2007ik,Boyle:2009vi},
and also has been observed in the black hole perturbation
approach~\cite{Sundararajan:2007jg,Burko:2010au}.
This above analysis is also applicable to the Weyl scalar.
In the next subsection,  we extend the perturbative formula to order $1/r^2$.

\subsection{Second order corrections: $(1/r^2)$-terms}\label{sec:2nd}

In this subsection, we discuss the next order correction of $\psi_4^{\ell m}$
in the $1/r$-expansion, first on a Schwarzschild background.
The starting point is the RWZ formalism
and Eq.~(\ref{eq:asymtRWZ}) is extended to order $1/r^2$.
For the even parity function, we have
\bea
\Psi_{\ell m}^{\rm (even)} &=& H^{\rm (even)}_{\ell m} (t-r^*) 
+ \frac{\ell(\ell+1)}{2\,r} \int dt H^{\rm (even)}_{\ell m} (t-r^*)
+ \frac{(\ell-1)\ell(\ell+1)(\ell+2)}{8\,r^2} \int \int dt \,dt H^{\rm (even)}_{\ell m} (t-r^*)
\cr &&
- \frac{3\,(\ell^2+\ell+2)\,M}{2\,(\ell-1)(\ell+2)\,r^2} \int dt H^{\rm (even)}_{\ell m} (t-r^*)
+ {\cal O}(1/r^3) \,,
\label{eq:asymt_even}
\eea
and for the odd parity function, 
\bea
\Psi_{\ell m}^{\rm (odd)} &=& H^{\rm (odd)}_{\ell m} (t-r^*) 
+ \frac{\ell(\ell+1)}{2\,r} \int dt H^{\rm (odd)}_{\ell m} (t-r^*)
+ \frac{(\ell-1)\ell(\ell+1)(\ell+2)}{8\,r^2} \int \int dt \,dt H^{\rm (odd)}_{\ell m} (t-r^*)
\cr &&
- \frac{3\,M}{2\,r^2} \int dt H^{\rm (odd)}_{\ell m} (t-r^*)
+ {\cal O}(1/r^3) \,. 
\label{eq:asymt_odd}
\eea
There is a difference between the even and odd parity functions at order $1/r^2$
due to the difference in the potentials of the RWZ equations.

Next, we convert the above even and odd parity functions into the Weyl scalar.
Using Eqs.~(C.1) and (C.2) in Ref.~\cite{Lousto:2005xu}
and taking care of the definitions in Eqs.~(\ref{eq:psi4_hp_hc}) and~(\ref{eq:RWZwaves}),
we obtain
\bea
\psi_{4\ell m}^{+} &=& \frac{1}{2\,r} \sqrt{\frac{(\ell+2)!}{(\ell-2)!}} 
\biggl[
\ddot{H}^{\rm (even)}_{\ell m} (t-r^*) 
+ \frac{(\ell-1)(\ell+2)}{2\,r} \dot{H}^{\rm (even)}_{\ell m} (t-r^*)
\cr && \qquad \qquad \qquad
+ \frac{(\ell-1)\ell(\ell+1)(\ell+2)}{8\,r^2} H^{\rm (even)}_{\ell m} (t-r^*)
+ \frac{3\,M}{2\,r^2} \dot{H}^{\rm (even)}_{\ell m} (t-r^*)
+ {\cal O}(1/r^3) \biggr] \,,
\cr
\psi_{4\ell m}^{-} &=& \frac{-i}{2\,r} \sqrt{\frac{(\ell+2)!}{(\ell-2)!}} 
\biggl[
\ddot{H}^{\rm (odd)}_{\ell m} (t-r^*) 
+ \frac{(\ell-1)(\ell+2)}{2\,r} \dot{H}^{\rm (odd)}_{\ell m} (t-r^*)
\cr && \qquad \qquad \qquad
+ \frac{(\ell-1)\ell(\ell+1)(\ell+2)}{8\,r^2} H^{\rm (odd)}_{\ell m} (t-r^*)
+ \frac{3\,M}{2\,r^2} \dot{H}^{\rm (odd)}_{\ell m} (t-r^*)
+ {\cal O}(1/r^3) \biggr] \,,
\eea
where the dot denotes the derivative with respect to the retarded time $(t-r^*)$.
The functions $\psi_{4\ell m}^{+/-}$ are defined in Eq.~(13) of Ref.~\cite{Lousto:2005xu}
and are the symmetric and antisymmetric Weyl scalar fields, respectively.
It is natural to have the same asymptotic behavior for the Weyl scalar fields.

Combining the above $\psi_{4\ell m}^{+/-}$
as $\psi_{4}^{\ell m}=\psi_{4\ell m}^{+} + \psi_{4\ell m}^{-}$, we obtain the extension of
Eq.~(\ref{eq:asymtPsi4}) as
\bea
r\,\psi_4^{\ell m} &=& \ddot {\tilde H}_{\ell m} (t-r^*) 
+ \frac{(\ell -1)(\ell +2)}{2\,r} \dot {\tilde H}_{\ell m} (t-r^*)
+ \frac{(\ell-1)\ell(\ell+1)(\ell+2)}{8\,r^2} {\tilde H}_{\ell m} (t-r^*)
\cr && 
+ \frac{3\,M}{2\,r^2} \dot{\tilde H}_{\ell m} (t-r^*)
+ {\cal O}(1/r^3) \,.
\label{eq:asymtPsi4_2nd}
\eea
Inverting this equation, the perturbative formula extended to order $1/r^2$
becomes
\bea
\left. r\,\psi_4^{\ell m}\right|_{r=\infty} &=& r\,\psi_4^{\ell m}(t,r)
- \frac{(\ell -1)(\ell +2)}{2\,r} \int dt \,[r\,\psi_4^{\ell m}(t,r)]
\cr &&
+ \frac{(\ell -1)(\ell +2)(\ell^2+\ell-4)}{8\,r^2} \int \int dt \,dt \,[r\,\psi_4^{\ell m}(t,r)]
- \frac{3\,M}{2\,r^2} \int dt \,[r\,\psi_4^{\ell m}(t,r)]
+ {\cal O}(1/r^3) \,.
\label{eq:asymtPsi4_2}
\eea

The above relation is  valid for the extrapolation of the $\psi_4$
in the Kinnersley tetrad. Next we will consider the corrections due to spin
and the use of a tetrad used in numerical relativity (NR) at a finite
$r$ and its decomposition into $(\ell, m)$ modes.

\section{In spinning background}\label{sec:PertSpin}

In this section, we include the spin dependence in the Teukolsky
formalism~\cite{Teukolsky:1973ha} of the Kerr (mass $M$ and Kerr parameter $a$)
black hole perturbation. 
It is noted that the wave function in the Teukolsky equation is
${}_{-2}\Psi=(r-i a \cos \theta)^4 \psi_4$.
Here, we ignore ${\cal O}(1/(\omega r)^3)$ and ${\cal O}((a\omega)^2)$
to derive a perturbative extrapolation formula from the frequency domain analysis.
For example, when the extraction radius is $r=100M$ for $a=M$,
the rough error estimation gives $1/(\omega r)^{3}=0.237\%$ and $(a\omega)^2=0.563\%$
at $\omega=0.075/M$, respectively.
This frequency is a reference 
to produce a hybrid post-Newtonian (PN)-NR waveform
for the $(\ell,m)=(2,2)$ mode
in the Numerical INJection Analysis (NINJA) project~\cite{Ajith:2012az}.

\subsection{Background spin correction}\label{sec:Kerr1}

First, we focus on the Teukolsky's wave function.
\bea
{}_{-2} \Psi = \int d\omega \sum_{\ell m} {}_{-2} \Psi_{\ell m \omega}(r)
{}_{-2} S_{\ell m}^{a\omega} (\theta,\phi) e^{-i \omega t} \,.
\label{eq:Teuk_full}
\eea
It is noted that we have used the spin-weighted spheroidal harmonics
(${}_{-2} S_{\ell m}^{a\omega}(\theta,\phi)$)
in the Teukolsky formalism, while the spin-weighted spherical harmonics
${}_{-2}Y_{\ell m}$ are used in the NR simulations.
The spin-weighted spheroidal harmonics,
which are the solution of the angular Teukolsky equation,
can be expanded as~\cite{Tagoshi:1996gh}
\bea
{}_{-2} S_{\ell m}^{a\omega} = {}_{-2}Y_{\ell m}
+ a\omega \sum_{\ell'} c_{\ell m}^{\ell'} {}_{-2}Y_{\ell' m} + {\cal O}((a\omega)^2) \,,
\eea
where the coefficient $c_{\ell m}^{\ell'}$ has a non-zero value only for $\ell'=\ell \pm 1$,
\bea
c_{\ell m}^{\ell-1} = -\frac{2}{\ell^2} \sqrt{\frac{(\ell+2)(\ell-2)(\ell+m)(\ell-m)}{(2\ell-1)(2\ell+1)}} \,,
\quad
c_{\ell m}^{\ell+1} = \frac{2}{(\ell+1)^2} \sqrt{\frac{(\ell+3)(\ell-1)(\ell+m+1)(\ell-m+1)}{(2\ell+1)(2\ell+3)}} \,.
\eea
The radial Teukolsky equation in the frequency domain gives the asymptotic solution,
\bea
\frac{{}_{-2}\Psi_{\ell m \omega}(r)}{r^3} &=&
\biggl[1+
\left( {\frac {-4\,ima}{\ell \left( \ell+1 \right)}}
+ {\frac {i \left( \ell-1 \right)\left( \ell+2 \right) }{2 \omega}} \right) \frac{1}{r}
\cr &&
+ \left(  \left( iM+{\frac { \left( \ell+2 \right)  \left( \ell-1 \right) }{\omega\,\ell \left( \ell+1 \right) }} \right) m a
+{\frac {3\,iM}{2\omega}}-\frac{1}{8}{\frac {\ell \left( \ell-1 \right)  \left( \ell+2 \right)  \left( \ell+1 \right) }{{\omega}^{2}}} \right)\frac{1}{r^2}
+ {\cal O}(1/(\omega r)^{3},(a\omega)^2 ) \biggr] H_{\ell m\omega} 
\cr
&=&
\biggl[1+ 
\left( {\frac {-4\,ima}{\ell \left( \ell+1 \right) }}
+ {\frac {i \left( \ell-1 \right)\left( \ell+2 \right) }{2 \omega}} \right) \frac{1}{r}
- \frac{1}{8}{\frac {\ell \left( \ell-1 \right)  \left( \ell+2 \right)  \left( \ell+1 \right) }{{\omega}^{2}}} \frac{1}{r^2}
+ \textrm{[higher}\,\,\textrm{order]} \biggr] H_{\ell m\omega} 
\,.
\label{eq:higherorderIG}
\eea
$H_{\ell m\omega}$ is related to the waveform at infinity.
In the last line of the above equation, we have ignored various cross terms which are included
in [higher order], 
$(M\omega)(a\omega)/(\omega r)^2$, $(a\omega)/(\omega r)^2$
and $(M\omega)/(\omega r)^2$
where we assumed that $M$ and $a$ are the same order.

Inserting ${}_{-2}\Psi_{\ell m \omega}(r)$ and ${}_{-2} S_{\ell m}^{a\omega}$
into Eq.~(\ref{eq:Teuk_full}), we have
\bea
{}_{-2} \Psi 
&=&
\int d\omega \sum_{\ell m} 
\biggl[
\left(1+ 
\left( {\frac {-4\,ima}{\ell \left( \ell+1 \right) }}
+ {\frac {i \left( \ell-1 \right)\left( \ell+2 \right) }{2 \omega}} \right) \frac{1}{r}
- \frac{1}{8}{\frac {\ell \left( \ell-1 \right)  \left( \ell+2 \right)  \left( \ell+1 \right) }{{\omega}^{2}}} \frac{1}{r^2}
\right) H_{\ell m\omega} {}_{-2} Y_{\ell m}
\cr && +
a\omega \left( c_{\ell m}^{\ell-1} {}_{-2}Y_{\ell-1 m} + c_{\ell m}^{\ell+1} {}_{-2}Y_{\ell+1 m} \right) H_{\ell m\omega} \biggr] e^{-i \omega t} 
+ \textrm{[higher}\,\,\textrm{order]}
\,.
\eea
Therefore, the spin-weighted spherical harmonic expansion becomes
\bea
\frac{{}_{-2}\Psi_{\ell m\omega}}{r^3} &=&
\int d\omega 
\biggl[
\left(1+ 
\left( {\frac {-4\,ima}{\ell \left( \ell+1 \right) }}
+ {\frac {i \left( \ell-1 \right)\left( \ell+2 \right) }{2 \omega}} \right) \frac{1}{r}
- \frac{1}{8}{\frac {\ell \left( \ell-1 \right)  \left( \ell+2 \right)  \left( \ell+1 \right) }{{\omega}^{2}}} \frac{1}{r^2}
\right) H_{\ell m\omega} 
\cr && +
a\omega \left( c_{\ell+1 m}^{\ell} H_{\ell+1 m\omega} + c_{\ell-1 m}^{\ell} H_{\ell-1 m\omega} \right) \biggr] e^{-i \omega t} 
+ \textrm{[higher}\,\,\textrm{order]}
\,.
\label{eq:improved_eq1}
\eea

\subsection{Use of the full numerical tetrad}\label{sec:Lazarus}

Eq.~(\ref{eq:Schw1}) in the nonspinning case relates the Weyl scalar $\psi_4$ at
a finite radii with the scalar at infinity. The preferred tetrad in
perturbation theory of black holes is the Kinnersley tetrad~\cite{Kinnersley_1969}
that make use of the algebraic specialty of the Kerr background where
$\psi_4$ vanishes. On the other hand, in full numerical relativity,
the lack of a reference background makes this choice ambiguous and
another tetrad, labeled `NR', is conventionally used. 
This variant of the `psikadelia' tetrad is described in Ref.~\cite{Baker:2001sf}.

Using Eq.~(2.15) in Ref.~\cite{Campanelli:2005ia},
we check the tetrad dependence.
Assuming the peeling theorem ($\psi_4 = [r\psi_4]/r,\,\psi_3 = [r^2\psi_3]/r^2,\,\psi_2 = [r^3\psi_2]/r^3,\,
\psi_1 = [r^4\psi_1]/r^4,\,\psi_0 = [r^5\psi_0]/r^5$,
where the functions in the square bracket are order $r^0$ for large $r$),
we have
\bea
r\psi_4^{\rm Kin} &=& 
\frac{1}{2}\,[r\psi_4^{\rm NR}]-{\frac {M [r\psi_4^{\rm NR}]}{{r}}}
-\frac{1}{4}\,{\frac {a \left( 7\,a [r\psi_4^{\rm NR}] \, \cos^{2} \theta
-3\,a [r\psi_4^{\rm NR}]  \right) }{{r}^{2}}}
\nonumber \\ &&
+i \left( {\frac {a\cos \theta [r\psi_4^{\rm NR}]}{{r}}}
-\frac{1}{4}\,{\frac {a \left( 4\,\sin \theta [r^2\psi_3^{\rm NR}]
+8\,[r\psi_4^{\rm NR}] \,\cos \theta M \right) }{{r}^{2}}} \right) + {\cal O}(1/r^3) \,.
\eea

After recasting the relationship between  ${}_{-2}\Psi$ and $\psi_4$
in terms of the NR $\psi_4$, we get
\bea
{}_{-2}\Psi &=& 
\frac{1}{2}\,{r}^{3}[r\psi_4^{\rm NR}]
- \left( M+i a \cos\theta \right) [r\psi_4^{\rm NR}]\,{r}^{2}
+ \left( 2\,iMa \cos \theta  [r\psi_4^{\rm NR}] -i a \sin \theta [r^2\psi_3^{\rm NR}] \right) r
+ {\cal O}(1/(\omega r)^0,(a\omega)^2)
\cr
&=&
\frac{1}{2}\,{r}^{3}[r\psi_4^{\rm NR}]
- \left( M+i a \cos\theta \right) [r\psi_4^{\rm NR}]\,{r}^{2}
+ \textrm{[higher}\,\,\textrm{order]}
\,,
\eea
where we have ignored various cross terms again.
The spin-weighted spherical harmonics expansion then becomes
\bea
\frac{{}_{-2}\Psi_{\ell m}}{r^3} &=& 
\left(\frac{1}{2} - \frac{M}{r} \right) [r\psi_{4\ell m}^{\rm NR}]
-\frac{i a}{r} \sum_{\ell' m'} C_{\ell m}^{\ell' m'} [r\psi_{4\ell' m'}^{\rm NR}] 
+ \textrm{[higher}\,\,\textrm{order]}
\,,
\label{eq:improved_eq2}
\eea
where $C_{\ell m}^{\ell' m'}$ is defined as
\bea
C_{\ell m}^{\ell' m'} = 
\int d\Omega {}_{-2}Y_{\ell m}^*(\Omega) \cos \theta {}_{-2}Y_{\ell' m'}(\Omega) \,,
\label{eq:Clmlm}
\eea
and has a non-zero values for $\ell'=\ell$ and $\ell'=\ell \pm 1$ with
$m'=m$ given by
\bea
C_{\ell m}^{\ell m} = \frac{2m}{\ell (\ell+1)} \,,
\quad
C_{\ell m}^{\ell+1\, m} =
\frac{1}{\ell+1}
\sqrt {{\frac {  \left( \ell-1 \right) \left( \ell+3 \right)  \left( \ell-m+1 \right)  \left( \ell+m+1 \right)  }
{  \left( 2\,\ell+1 \right) \left( 2\,\ell+3 \right) }}} \,,
\eea
(see also Appendix A of Ref.~\cite{Berti:2014fga}). 
Because of the above result, we may consider $C_{\ell m}^{\ell' m'}=C_{\ell' m'}^{\ell m}$.

\subsection{Improved extrapolation formula}\label{sec:Improved}

Comparing Eqs.~(\ref{eq:improved_eq1}) and~(\ref{eq:improved_eq2})
in the time domain, we have
\bea
\ddot {\tilde H}_{\ell m} (t-r^*) 
&&
+ \frac{\left( \ell-1 \right)  \left( \ell+2 \right)}{2\,r} \dot {\tilde H}_{\ell m} (t-r^*)
- {\frac {4\,i\, ma}{\ell \left( \ell+1 \right)\,r }} \ddot {\tilde H}_{\ell m} (t-r^*)
+ \frac{\ell \left( \ell-1 \right)  \left( \ell+2 \right)  \left( \ell+1 \right)}{8\,r^2} {\tilde H}_{\ell m} (t-r^*)
\cr && 
+ i\,a \left( c_{\ell+1 m}^{\ell} \dddot {\tilde H}_{\ell+1 m} (t-r^*) + c_{\ell-1 m}^{\ell} \dddot {\tilde H}_{\ell-1 m} (t-r^*) \right)
\cr &&
=
\left(\frac{1}{2} - \frac{M}{r} \right) [r\psi_{4\ell m}^{\rm NR}]
-\frac{i a}{r} \sum_{\ell' m'} C_{\ell m}^{\ell' m'} [r\psi_{4\ell' m'}^{\rm NR}] 
+ \textrm{[higher}\,\,\textrm{order]} \,.
\eea
Our improved extrapolation formula derived from the above equation is
therefore
\bea
\left. r\,\psi_4^{\ell m}\right|_{r=\infty}
&=&
\left(1-{\frac {2M}{{r}}}\right) \left( r\psi_{4\ell m}^{\rm NR}(t,r) 
- \frac{(\ell -1)(\ell +2)}{2\,r} \int dt [r\psi_{4\ell m}^{\rm NR}(t,r)]
\right. \cr && \qquad \left.
+ \frac{(\ell -1)(\ell +2)(\ell^2+\ell-4)}{8\,r^2} \int \int dt dt [r\psi_{4\ell m}^{\rm NR}(t,r)]
\right)
\cr &&
+ \frac{2\,i\,a}{(\ell+1)^2} \sqrt{\frac{(\ell+3)(\ell-1)(\ell+m+1)(\ell-m+1)}{(2\ell+1)(2\ell+3)}}
\left( [r\partial_t{\psi}_{4\ell+1 m}^{\rm NR}(t,r)]
- \frac{\ell (\ell+3)}{r} [r\psi_{4\ell+1 m}^{\rm NR}(t,r)] \right)
\cr &&
- \frac{2\,i\,a}{\ell^2} \sqrt{\frac{(\ell+2)(\ell-2)(\ell+m)(\ell-m)}{(2\ell-1)(2\ell+1)}}
\left( [r\partial_t{\psi}_{4\ell-1 m}^{\rm NR}(t,r)]
- \frac{(\ell-2)(\ell+1)}{r} [r\psi_{4\ell-1 m}^{\rm NR}(t,r)] \right)
\cr &&
+ \textrm{[higher}\,\,\textrm{order]} 
\,.
\label{eq:improved_formula}
\eea

The above formula (\ref{eq:improved_formula}) is our definitive equation for
extrapolation of the waveform at finite radii to order $1/r^2$.
It involves the first order correction in the mass (Schwarzschild-like)
and spin (Kerr-like) of the sources, and corrects for the differences
between the numerical and Kinnersley tetrads.

On the other hand, we have proposed an extrapolation formula
to order $1/r$ in~\cite{Nakano:2015rda}
\bea
\label{eq:definitive1}
\left. r\,\psi_4^{\ell m}\right|_{r=\infty}
&=&
\left(1-{\frac {2M}{{r}}}\right) \left( r\psi_{4\ell m}^{\rm NR}(t,r) 
- \frac{(\ell -1)(\ell +2)}{2\,r} \int dt [r\psi_{4\ell m}^{\rm NR}(t,r)] 
\right)
\cr &&
- \frac{2\,i\,a}{r} 
\sum_{\ell' \neq \ell,\, m'=m} [r\psi_{4\ell' m'}^{\rm NR}(t,r)] 
C_{\ell m}^{\ell' m'} \,.
\eea
Since we did not take care of the difference
between the spin-weighted spheroidal and spherical harmonics
in the above equation, 
the spin correction is different between 
Eqs.~(\ref{eq:improved_formula}) and~(\ref{eq:definitive1}).

In the equations above $r$ is the areal radius (Schwarzschild coordinate
in the nonrotating case). In the standard numerical simulations we use
$R_{\rm NR}$ that asymptotically behaves more like, $R$, the 
`isotropic' radial coordinate, hence in Eq.~(\ref{eq:improved_formula}) 
we typically use $r=R\left(1+(M+a)/(2R)\right)\left(1+(M-a)/(2R)\right)$. 
Alternatively, one could also compute directly the areal radius from
the full numerical simulation via $r=\sqrt{A(R)/4\pi}$, where $A(R)$
is the measure surface area of the `sphere' $R={\rm const}$.

\section{Estimation of the radiated energy and momenta}\label{sec:Lazarus_E}

Using the improved extrapolation formula in Eq.~(\ref{eq:improved_formula}),
we derive extrapolation formulas for the radiated energy and momenta
which are calculated from the Weyl scalar $\psi_4$ as~\cite{Campanelli:1998jv}
\bea
\frac{dE}{dt} &=& \frac{1}{16\pi}
\int d\Omega \left| \int dt \left. r \psi_4\right|_{r=\infty} \right|^2 
\cr
&=& \frac{1}{16\pi} \sum_{\ell m} \left| \int dt \left. r \psi_{4\ell m}\right|_{r=\infty} \right|^2 
\,,
\cr
\frac{dL_z}{dt} &=& - \frac{1}{16\pi}
\Re \left[ \int d\Omega \partial_{\phi} \left( \int dt \left. r \psi_4\right|_{r=\infty} \right)
\left( \int \int dt dt \left. r \bar{\psi}_4\right|_{r=\infty} \right) \right]
\cr
&=& \frac{1}{16\pi} \Im \sum_{\ell m} m
\left( \int dt \left. r \psi_{4\ell m}\right|_{r=\infty} \right)
\left( \int \int dt dt \left. r \bar{\psi}_{4\ell m}\right|_{r=\infty} \right)
\,,
\cr
\frac{dP_z}{dt} &=& \frac{1}{16\pi}
\int d\Omega \cos \theta \left| \int dt \left. r \psi_4\right|_{r=\infty} \right|^2 
\cr
&=&
\frac{1}{16\pi} \sum_{\ell m}\sum_{\ell' m'} C_{\ell m}^{\ell' m'}
\left(\int dt \left. r \psi_{4\ell m}\right|_{r=\infty} \right) 
\left(\int dt \left. r \bar{\psi}_{4\ell' m'}\right|_{r=\infty} \right) \,.
\eea
Here, we focus only on the $z$ component for the angular and linear momenta,
and have used the normalization of $\psi_4$ as Eq.~(\ref{eq:psi4_hp_hc}).
$C_{\ell m}^{\ell' m'}$ is the same as in Eq.~(\ref{eq:Clmlm}).

Inserting Eq.~(\ref{eq:improved_formula}) into the above expressions,
we obtain the extrapolation formulas
\bea
\frac{dE}{dt}
&=& \frac{1}{16\pi} \sum_{\ell m}
\biggl[
\left(1- \frac{4M}{r} \right) \left| \int dt \Phi_{\ell m} \right|^2
+ \frac{(\ell-1)(\ell+2)}{2r^2} 
\left| \int \int dt dt \Phi_{\ell m} \right|^2
\biggr)
\cr &&
-4 a \biggl( \frac{C_{\ell m}^{\ell+1 m}}{\ell+1} 
\Im \left(\Phi_{\ell+1 m} \left(\int dt \bar{\Phi}_{\ell m}\right) \right)
- \frac{C_{\ell m}^{\ell-1 m}}{\ell} 
\Im \left(\Phi_{\ell-1 m} \left(\int dt \bar{\Phi}_{\ell m}\right) \right)
\biggr)\biggr] \,,
\cr
\frac{dL_z}{dt}
&=& \frac{1}{16\pi} \sum_{\ell m} m
\biggl[
\left(1- \frac{4M}{r} \right) \Im \left(\left(\int dt \Phi_{\ell m}\right) \left( \int \int dt dt \bar{\Phi}_{\ell m} \right) \right)
\cr &&
+ \frac{(\ell-1)(\ell+2)}{2r^2} 
\Im \left(
\left(\int \int dt dt \Phi_{\ell m}\right) \left(\int \int \int dt dt dt \bar{\Phi}_{\ell m}\right) \right)
\cr &&
-4 a \biggl( \frac{C_{\ell m}^{\ell+1 m}}{\ell+1} 
\Re \left(\left(\int dt \Phi_{\ell+1 m}\right) \left(\int dt \bar{\Phi}_{\ell m}\right) \right)
- \frac{C_{\ell m}^{\ell-1 m}}{\ell} 
\Re \left(\left(\int dt \Phi_{\ell-1 m}\right) \left(\int dt \bar{\Phi}_{\ell m}\right) \right)
\biggr)
\biggr] \,,
\cr
\frac{dP_z}{dt} 
&=& \frac{1}{16\pi} \sum_{\ell m} \sum_{\ell' m'}
C_{\ell m}^{\ell' m'}
\biggl[
\left(1- \frac{4M}{r} \right) \left(\int dt \Phi_{\ell m}\right) \left( \int dt \bar{\Phi}_{\ell' m'} \right) 
\cr &&
- \frac{(\ell-\ell')(\ell+\ell'+1)}{2 r}
\left(\int dt \Phi_{\ell m}\right) 
\left(\int \int dt dt \bar{\Phi}_{\ell' m'}\right)
\cr &&
- \frac{2 (\ell-1)\ell(\ell+1)(\ell+2)
- \ell(\ell+1)\ell'(\ell'+1)+4}{4r^2}
\Re \left(
\left(\int \int dt dt \Phi_{\ell m}\right)
\left(\int \int dt dt \bar{\Phi}_{\ell' m'}\right) \right)
\cr &&
- 4 a \biggl( \frac{C_{\ell m}^{\ell+1 m}}{\ell+1} 
\Im \left( \Phi_{\ell+1 m} \left(\int dt \bar{\Phi}_{\ell' m'}\right) \right)
- \frac{C_{\ell m}^{\ell-1 m}}{\ell} 
\Im \left( \Phi_{\ell-1 m} \left(\int dt \bar{\Phi}_{\ell' m'}\right) \right)
\biggr)
\cr &&
+ \frac{2 a}{r} \biggl(
\frac{(2\ell (\ell+3) -(\ell'-1)(\ell'+2)) C_{\ell m}^{\ell+1 m}}{\ell+1}
\Re \left( \left(\int dt \Phi_{\ell+1 m}\right)
\left(\int dt \bar{\Phi}_{\ell' m'}\right) \right)
\cr && \qquad
- \frac{(2(\ell-2) (\ell+1) -(\ell'-1)(\ell'+2)) C_{\ell m}^{\ell-1 m}}{\ell}
\Re \left( \left(\int dt \Phi_{\ell-1 m}\right)
\left(\int dt \bar{\Phi}_{\ell' m'}\right) \right)
\biggr)
\biggr] \,,
\eea
where $\Phi_{\ell m} = r\psi_{4\ell m}^{\rm NR}(t,r)$
and we have ignored [higher order] terms described below Eq.~(\ref{eq:higherorderIG}).
In order to simplify the expressions and
to reduce the order of integration with respect to time,
we have used the frequency domain analysis.
In the expression of the radiated linear momentum,
we take the sum over $\ell'$ and $m'$ as $\ell'=\ell,\,\ell\pm 1$ and $m'=m$.
It should be noted that the ($\ell,\, m$) mode denotes
the index of the spin-weighted spherical harmonics.
There are order $1/r$ corrections for the radiated energy and angular momentum
that is different from Ref.~\cite{Nakano:2015rda} because of the tetrad difference.

\section{Full numerical implementation}\label{sec:FN}

In order to evaluate the actual benefits of the analytic expression
(\ref{eq:definitive1}) 
for the extrapolation to infinity of gravitational waveforms
extracted at a finite radii in a typical full numerical setting we
consider the test case  described in Table \ref{tab:appID}.
We perform three sets of runs with increasing global resolution in
the convergence regime and we extract waveforms at ten different
radii, evenly  separated as $1/r$.

\begin{table}[!ht]
\caption{Initial data for our test case. The binary's parameters were
estimated using  quasicircular orbits.}\label{tab:appID}
\begin{ruledtabular}
\begin{tabular}{lcccccccccccc}
Config.   & $x_1/M$ & $x_2/M$  & $P/M$    & $m^p_1/M$ & $m^p_2/M$ & $S_1/M^2$ & $S_2/M^2$ & $m^H_1/M$ & $m^H_2/M$ & $M_{\rm ADM}/M$ & $a_1/m_1^H$ & $a_2/m_2^H$\\
\hline
A\_DU0.8  & -4.9832 & 4.5267 & 0.09905 & 0.30178 & 0.30168 & -0.2     & 0.2     & 0.5    & 0.5    & 0.98951 & -0.8 & 0.8 \\
\end{tabular}
\end{ruledtabular}
\end{table}

In this work, we use a grid structure with 10 levels of refinement. 
The outer boundary was placed at 400M and for the medium resolution run
the resolution was $4M$ on the coarsest level and $1M$ in the wavezone.
The finest level around each BH was as wide as twice the
diameter of the relaxed horizon.
We also performed a lower and higher resolution run with resolutions in
the wavezone of $M/0.88$ and $M/1.2$.

The simulation results will depend on the extraction radii as well as on the
truncation errors due to finite resolution.
Hence we consider different resolutions and extraction radii and extrapolations
to null infinity.
In this paper we used extraction radii up to
$R_{\rm obs}/M=190$ and locate the extraction radii
equidistant in $1/R$, with $R_{\rm obs}/M=75,80.4,86.7,94.0,102.6,113.0,125.7,141.7,162.3,190.0.$ 

\begin{figure}[!ht]
  \includegraphics[width=0.32\columnwidth]{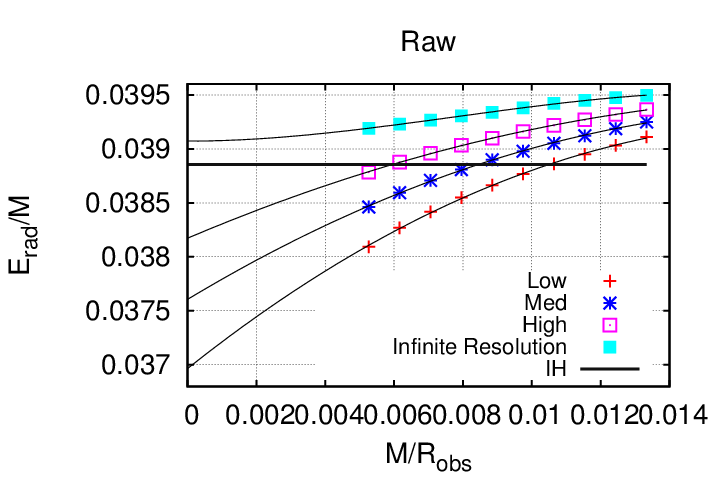}
  \includegraphics[width=0.32\columnwidth]{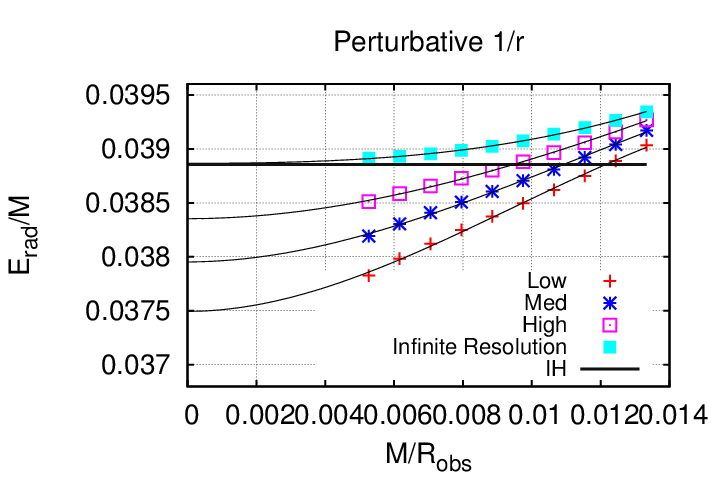}
  \includegraphics[width=0.32\columnwidth]{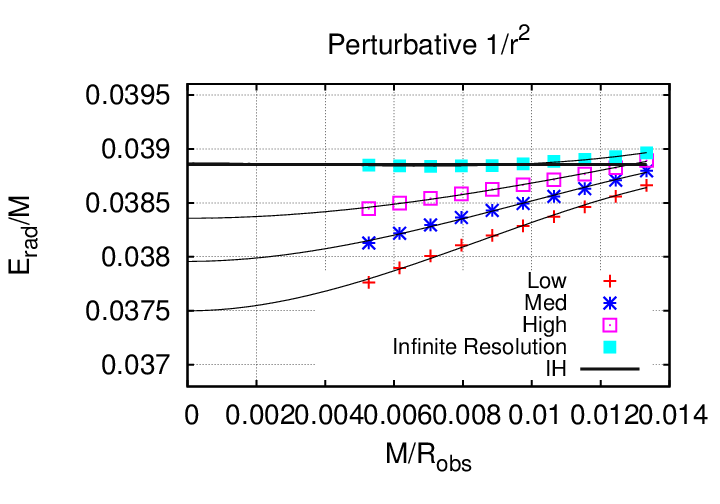}
  \caption{The energy radiated (adding up to $\ell=6$)
as a function of the observer location
$M/R_{\rm obs}=1/75$, $1/80.4$, $1/86.7$, $1/94.0$, $1/102.6$, $1/113.0$, $1/125.7$, $1/141.7$, $1/162.3$, $1/190.0$
for the directly extracted waveform, labeled as
`Raw' (left) and for the analytically extrapolated waveform, labeled
as `Perturbative $1/r$' (center) and `Perturbative $1/r^2$' (right).
}  \label{fig:Erad}
\end{figure}

\begin{figure}[!ht]
  \includegraphics[width=0.32\columnwidth]{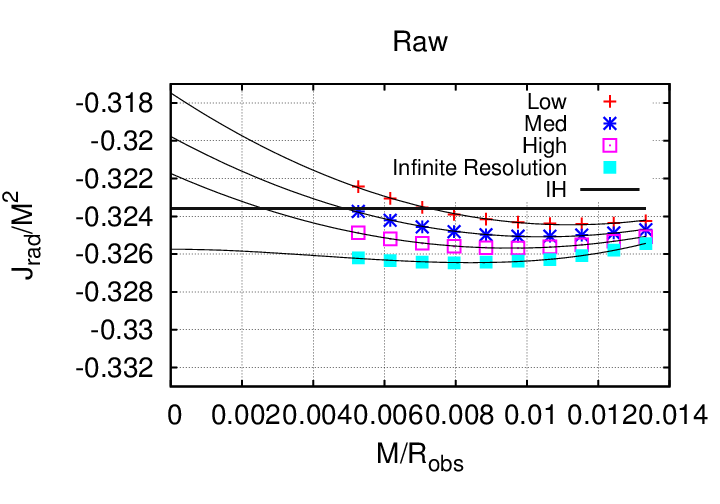}
  \includegraphics[width=0.32\columnwidth]{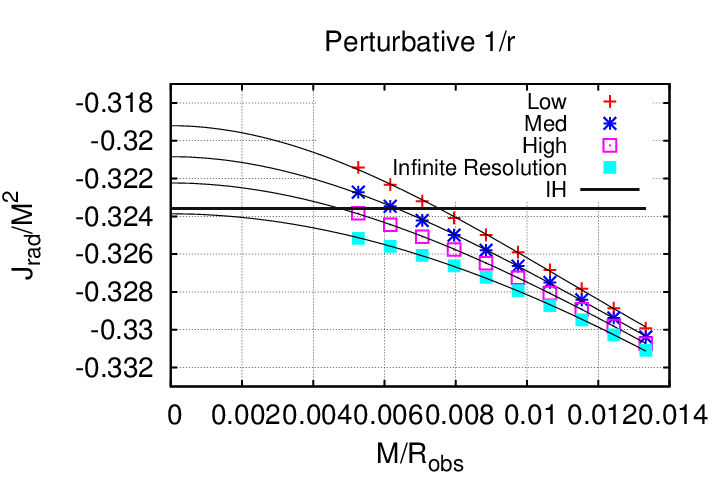}
  \includegraphics[width=0.32\columnwidth]{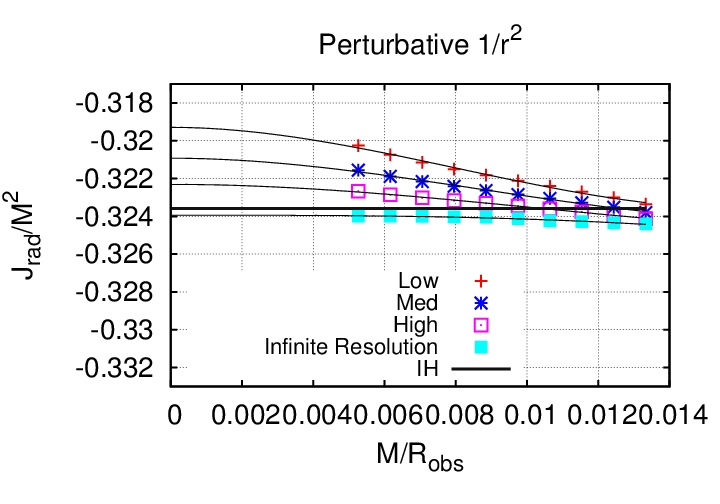}
  \caption{The Angular Momentum radiated (adding up to $\ell=6$)
as a function of the observer location
$M/R_{\rm obs}=1/75$, $1/80.4$, $1/86.7$, $1/94.0$, $1/102.6$, $1/113.0$, $1/125.7$, $1/141.7$, $1/162.3$, $1/190.0$
for the directly extracted waveform, labeled as
`Raw' (left) and for the analytically extrapolated waveform, labeled
as `Perturbative $1/r$' (center) and `Perturbative $1/r^2$' (right).
}
  \label{fig:Jrad}
\end{figure}

\begin{figure}[!ht]
  \includegraphics[width=0.32\columnwidth]{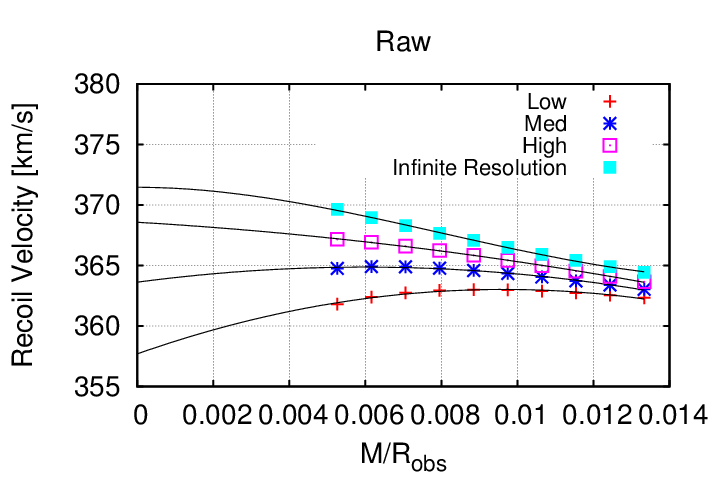}
  \includegraphics[width=0.32\columnwidth]{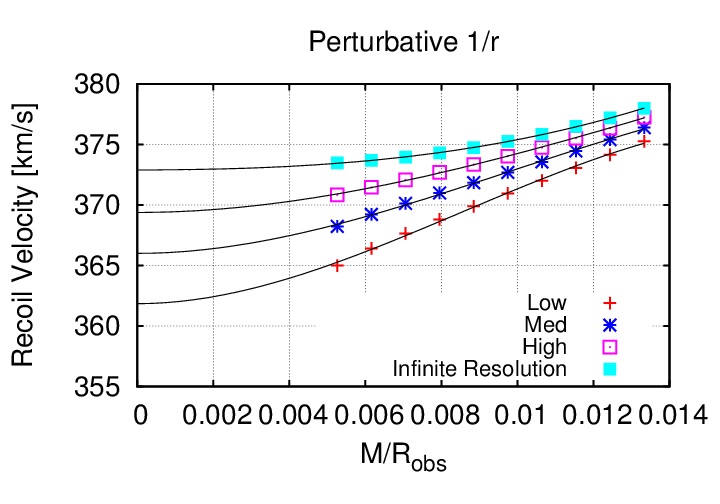}
  \includegraphics[width=0.32\columnwidth]{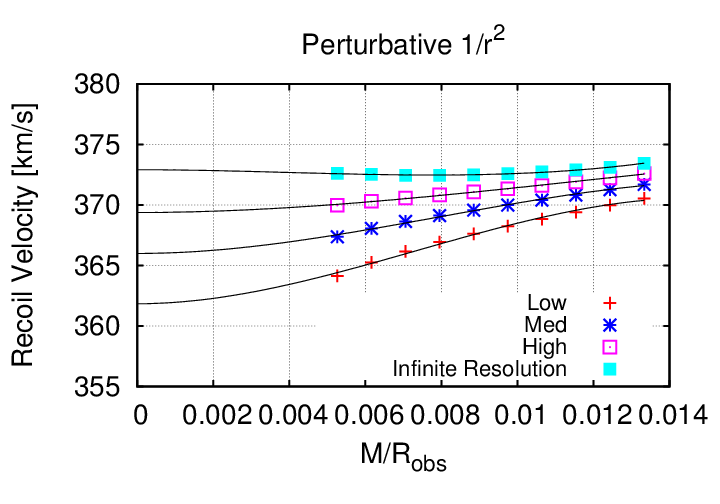}
  \caption{The Linear Momentum radiated (adding up to $\ell=6$)
as a function of the observer location
$M/R_{\rm obs}=1/75$, $1/80.4$, $1/86.7$, $1/94.0$, $1/102.6$, $1/113.0$, $1/125.7$, $1/141.7$, $1/162.3$, $1/190.0$
for the directly extracted waveform, labeled as
`Raw' (left) and for the analytically extrapolated waveform, labeled
as `Perturbative $1/r$' (center) and `Perturbative $1/r^2$' (right).
}
  \label{fig:Kick}
\end{figure}

We directly compared waveforms extracted with the characteristic
method to our extrapolation formula, Eq.~(\ref{eq:Schw1}), 
in Ref.~\cite{Babiuc:2010ze}, Figs.~8-9, 
and to purely numerical extrapolations in Ref.~\cite{Hinder:2013oqa}.
There we observed an excellent agreement with our analytic expression at
first order in $1/r$ for the phase, as predicted by the error analysis
of Eq.~(\ref{eq:PHcorrection}). 
The improvements in the amplitude are of higher order
as shown in  Eq.~(\ref{eq:APcorrection}).
In order to supplement those studies, here we focus on the integral
expressions for the energy and momenta radiated at infinity.
The results of such studies is displayed in 
Figs.~\ref{fig:Erad}-\ref{fig:Kick}. The radiated quantities are calculated 
using all modes up to $\ell=6$ where the news and strain are calculated via
the fixed frequency integration \cite{Campanelli:2008nk,Reisswig:2010di}.

In Fig.~\ref{fig:Erad} we observe the computed radiated energy directly
from the finite radii extraction that we denote as `Raw'. The figure
displays the different extraction radii, evenly distributed versus
$M/R_{obs}$ for the three finite-difference resolutions considered, denoted
as Low, Medium and High. We provide a Richardson extrapolation to infinite
resolution (3rd order) for each observer location value based on those
three resolutions and also the value of the total radiated energy as
inferred from the subtraction of the final horizon mass to
the initial ADM mass of the system
 (denoted by the thick straight line). This measure of the final black hole
mass, is very robust (at this scale) with increasing
resolution and provides a very accurate measure as well as a consistency
check of the extraction process.

We observe that for the `Raw' extraction increasing resolution
(particularly for the closer to sources observers) brings the results
further apart from the reference value inferred by the final horizon
mass. To get consistency, one needs to  first extrapolated to infinite
observer location and then to infinite resolution.

The second panel of Fig.~\ref{fig:Erad} displays the same computation
of the radiated energy, but after extrapolation of the waveforms
via Eq.~(\ref{eq:Schw1}). We use the extrapolation at each observer location.
We would expect that the dependence of the estimated energy radiated
with the observer location is weaker since we are correcting for the
$1/r$ behavior and only higher power dependencies should appear. We 
indeed observe flatter curves at all three finite-difference resolutions
for this case compared to the `Raw' extraction. The second feature is
that at a single observer location the values converge towards the
horizon value with increasing resolution. 
This is a desired feature, especially for a more demanding simulation
where one only has access to accurate extraction in the intermediate
zone between the sources and the radiation zone.
The third panel shows the extrapolation carried to
order $1/r^2$ using Eq.~(\ref{eq:improved_formula}) with $a=0$
(in practice we did not see a strong dependence on $a$).
Notably, in both cases, extrapolation to infinite resolution and
infinite observer location leads to values within $0.1\%$ of the
correct value as inferred by the horizon measure.

A similar behavior is observed in the computation of the angular
momentum radiated as displayed in Fig.~\ref{fig:Jrad}. For the
first panel with the `Raw' waveforms we see that increasing the
finite-difference resolution leads to extrapolated values further
apart from the horizon measure derived as the difference of
the final spin of the black hole \cite{Dreyer:2002mx}
to the initial total ADM angular
momentum (denoted by the thick straight line). Using the 
perturbative $1/r$ and $1/r^2$ extrapolations before the calculation 
of the angular momentum, as shown in the middle and left panels, 
leads to flatter curves with observer location and exhibits 
convergence toward the correct value with increasing finite-difference
resolution. In both cases, the extrapolation to both infinite resolution and
infinite observer location leads to predictions within $0.1\%$ of the
expected value. The importance of the extrapolation formula is that this
can also be achieved with information from a single finite observer location.

Finally we also compute the linear momentum radiated by the system
and display the results in Fig.~\ref{fig:Kick}. The first observation
is that we do not have a very accurate measure on the final horizon
for the recoil velocity to use as a reference value (although, see
the work of Ref.~\cite{Krishnan:2007pu}). 
However, based on the extrapolated values
we estimate the recoil velocity to lie in the range $372-373$km/s.
We then observe that at a given finite value of the observer, particularly
for those closer to the sources, the perturbative extrapolations
values lie closer to the expected recoil. The curves also look flatter
indicating the internal consistency of the extrapolation process.

\begin{figure}[!ht]
\vbox{
  \hbox{
    \includegraphics[width=0.32\columnwidth]{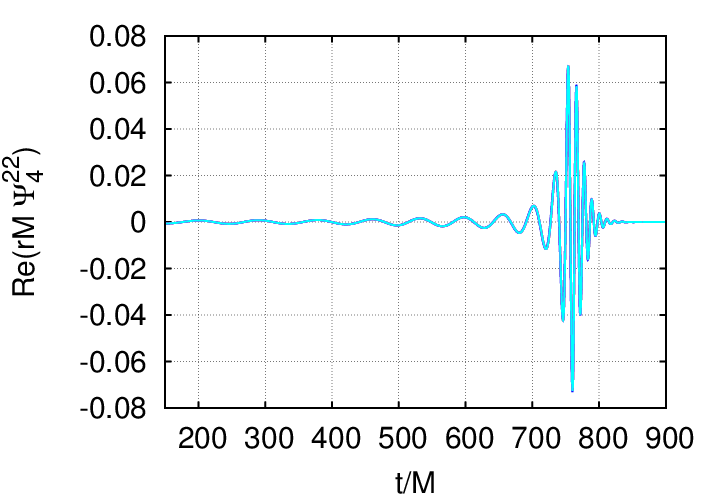}
    \includegraphics[width=0.32\columnwidth]{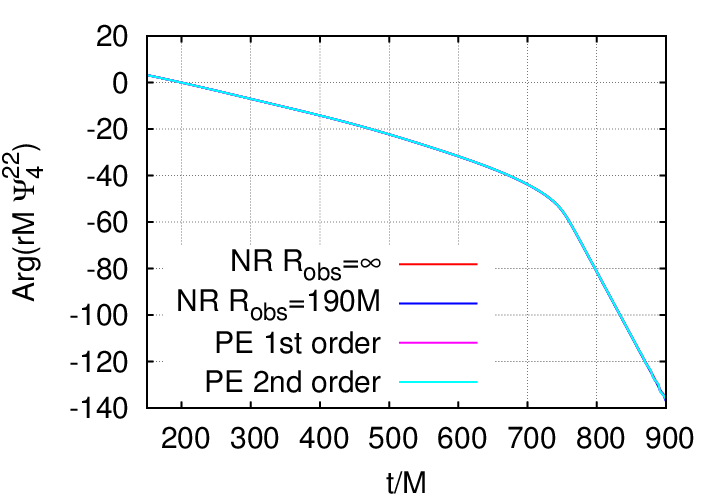}
    \includegraphics[width=0.32\columnwidth]{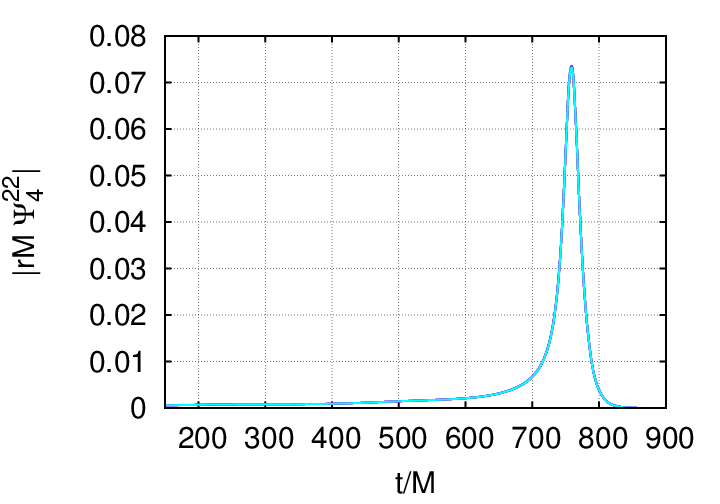}
  }
  \hbox{
    \includegraphics[width=0.32\columnwidth]{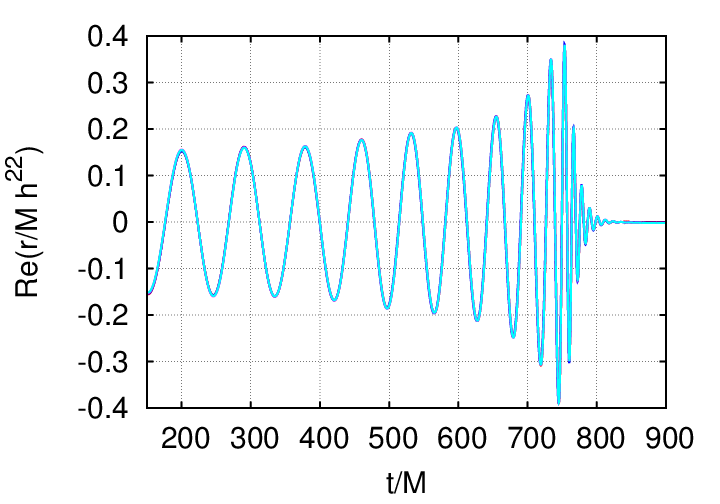}
    \includegraphics[width=0.32\columnwidth]{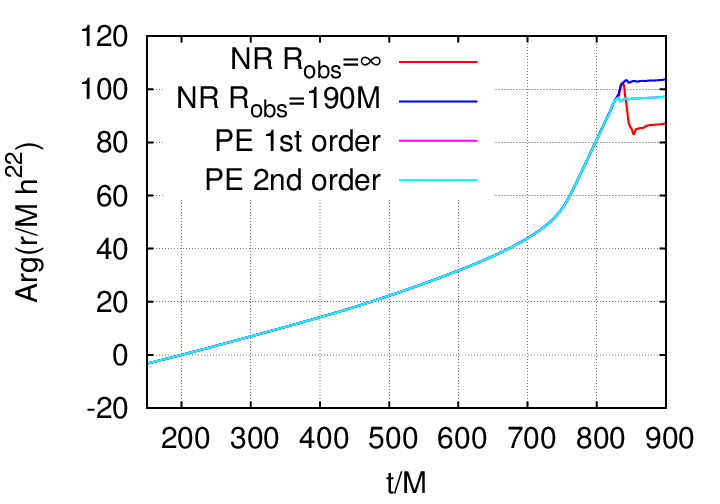}
    \includegraphics[width=0.32\columnwidth]{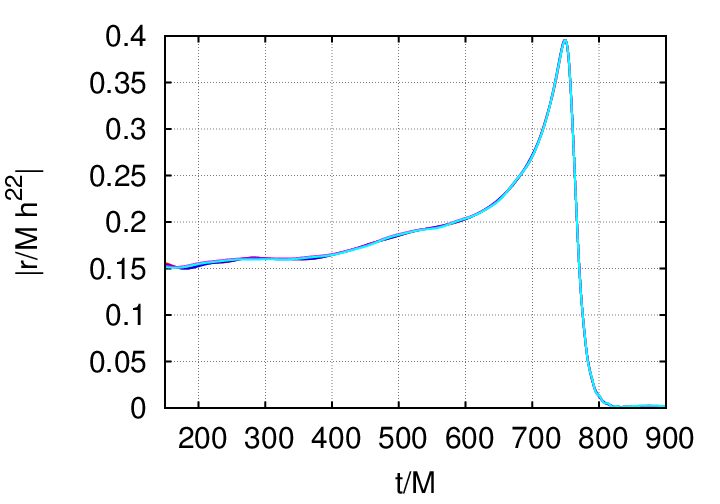}
  }
}
  \caption{`$\infty$' is the
extrapolated NR waveform (phase and amplitude) at highest resolution 
using 10 radii equally spaced between $75$ and $190M$. We
compare the above best extrapolated waveform, `$\infty$',
with that directly extracted at $R_{\rm obs}=190M$, 1st order, 
$(1/r)$, perturbative extrapolation 
(PE), and 2nd order,$(1/r^2)$, PE. In all cases the mode $(\ell,m)=(2,2)$ is displayed.  
The first row is the Weyl Scalar  $rM \Psi_4^{22}$ and the second row is the
gravitational strain $r h^{22}/M$.
}
  \label{fig:Amphase}
\end{figure}

\begin{figure}[!ht]
  \includegraphics[width=0.49\columnwidth]{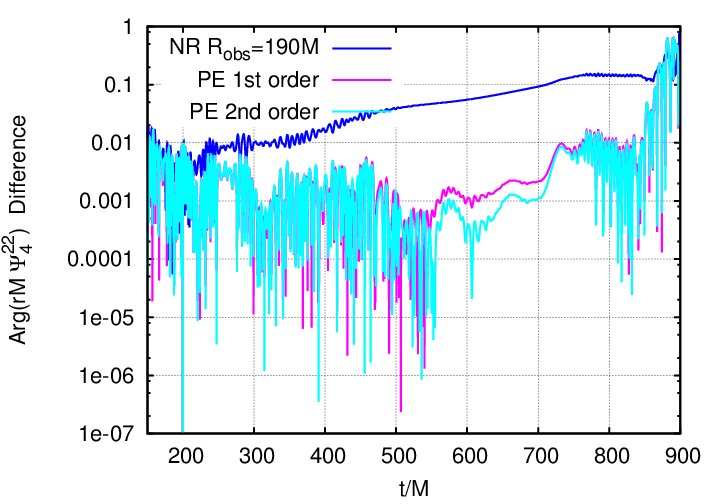}
  \includegraphics[width=0.49\columnwidth]{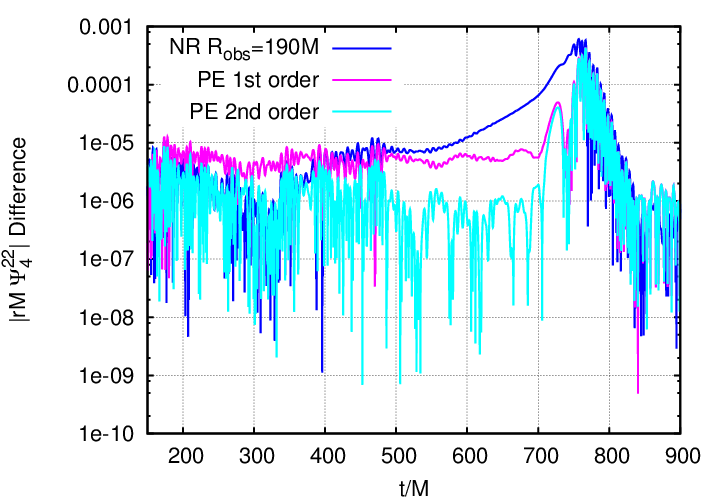}
  \caption{
Comparison of the best extrapolated waveform, `$\infty$',
with that directly extracted at $R_{\rm obs}=190M$, 1st order, $(1/r)$, 
perturbative extrapolation 
(PE), and 2nd order, $(1/r^2)$,  PE, calculating the difference abs(`$\infty$'-waveform) for
the phase and amplitude. The mode $(\ell,m)=(2,2)$ is displayed.
}
  \label{fig:DiffAmphase}
\end{figure}

\begin{figure}[!ht]
  \includegraphics[width=0.49\columnwidth]{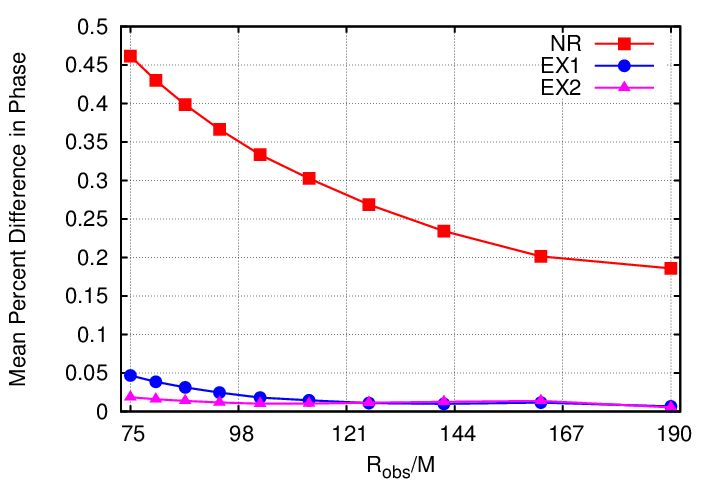}
  \includegraphics[width=0.49\columnwidth]{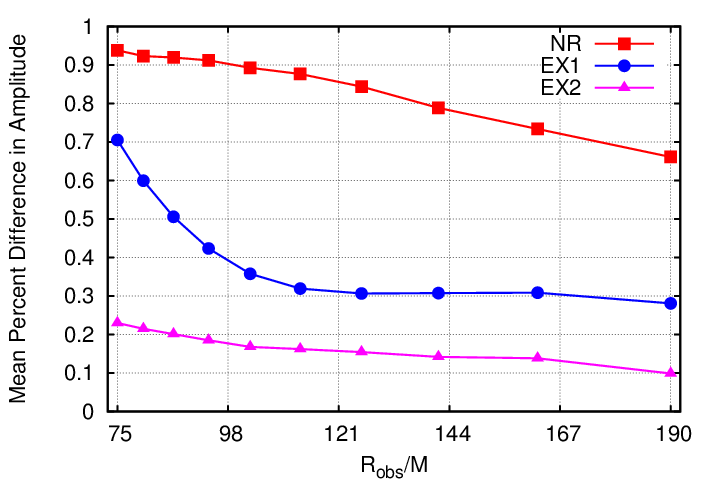}
  \caption{$(\ell,m)=(2,2)$ mode of the
raw (NR) and first (EX1) and second (EX2) perturbative extrapolated waveforms
as a function of radius versus the waveform extrapolated
to infinity.  Displayed is the mean of the $\%$ difference between
perturbative and extrapolated for each radii for the amplitude and phase
for the times between $400M$ and $800M$.
}
  \label{fig:secondAphi}
\end{figure}

In order to produce a reference waveform that we may consider the 
best extrapolation and hence approximation to the exact waveform in
Fig.~\ref{fig:Amphase}, we took the highest resolution run and used 
the ten extraction radii we
have to extrapolate the waveform in time using a 2nd order fitting
polynomial in $1/r$.  
We extrapolated the amplitude and phase after shifting the time by the
tortoise radius for each extraction radius.  We then can compare the
amplitude and phase of this extrapolated waveform to a finite radius
waveform ($R_{\rm obs}=190M$, our largest extraction radius), and to the waveforms
produced by using the $1/r$ and $1/r^2$ order perturbative extrapolations
(without the terms depending on the spin). The results are displayed
in Fig.~\ref{fig:DiffAmphase} which shows the benefits of using
our formulas  to approximate the waveform phase and amplitude
at infinity. Note that given the different dependence of the
phase correction ($1/r$ as shown in Eq.~(\ref{eq:PHcorrection})) and the
amplitude correction ($1/r^2$ as shown in Eq.~(\ref{eq:APcorrection}))
the phase and amplitude show further improvements by including the 
second order corrections. 
This is more explicitly displayed in 
Fig.~\ref{fig:secondAphi}, that summarized the averaged differences.

\section{Discussion}\label{sec:discussion}

In this paper we describe a procedure for extrapolating the waveform
at finite radius to infinity as a power series in $1/r$. We provided
the complete $1/r$ correction to the waveform in
 Eq.~(\ref{eq:definitive1}), including the spin terms in
the background. We have also found it important to include the
leading terms in $1/r^2$ in the extrapolation formula as given in
Eq.~(\ref{eq:improved_formula}).
We have tested the extrapolation formula's  properties in a typical
full numerical simulation of a black-hole binary,  where we can
verify the behavior of the extrapolation with different observer
locations and finite-difference resolutions. In numerical simulations
where the typical wavelength is relatively small compared to the
boundaries of the simulation, the perturbative extraction provides
at least a way of verifying the accuracy and consistency
of the waveforms and radiative quantities such as the total energy
and linear and angular momenta. 
In situations where it is extremely costly or inaccurate to 
extract at distances of two gravitational wavelengths from the
sources (rule of thumb for the radiation zone), this method provides
a crucial technique to evaluate waveforms and radiated quantities.
In particular, we have seen that
it is only the extrapolated waveform that converges with increasing
resolution to the correct values and that extrapolation to infinite
resolution of a finite extraction waveform can lead to a worse approximation.
Although, for far enough location observer and resolution these two
extrapolation processes eventually tend to commute.

The second order correction provided in Eq.~(\ref{eq:improved_formula})
could be useful in situations where we have extended sources or
one needs extreme resolutions near the sources and the simulation
grid cannot reach the radiation zone. It also provides an independent
way to estimate the errors of the first order extrapolation formula
 Eq.~(\ref{eq:definitive1}) by looking at the differences produced by
these two extrapolation formulas.

\acknowledgments 

The authors gratefully acknowledge the NSF for financial support from Grants
PHY-1305730, PHY-1212426, PHY-1229173,
AST-1028087, PHY-0969855, OCI-0832606, and
DRL-1136221. Computational resources were provided by XSEDE allocation
TG-PHY060027N, and by NewHorizons and BlueSky Clusters 
at Rochester Institute of Technology, which were supported
by NSF grant No. PHY-0722703, DMS-0820923, AST-1028087, and PHY-1229173.
H.N. acknowledges support by the Grant-in-Aid for Scientific Research
No.~24103006.

\appendix

\section{Second order correction with $\psi_0^{\ell m}$}\label{sec:2nd_c}

In Sub-Sec.~\ref{sec:2nd}, the formula has a term,
$\psi_4^{\ell m}$ integrated twice in time.
In order to remove this term, we may use the identities in the Teukolsky formalism,
and use the notation as given in Ref.~\cite{Keidl:2010pm}.
In the Schwarzschild background with mass $M$,
the Weyl scalar $\psi_4$ and $\psi_0$ 
have following relations.
\bea
r^4 \psi_4 = \frac{1}{32} r^4 f^2 \left(\frac{1}{f} \partial_t - \partial_r \right)^4
r^4 f^2 {\bar \Psi} \,,
\quad
\psi_0 = \frac{1}{8} \left( \edth^4 {\bar \Psi} + 12 M \partial_t \Psi \right) \,,
\eea
where $f=1-2M/r$ and $\edth=-(\partial_\theta-s \cot \theta + i \csc \theta \partial_\phi)$ for 
the spin-s weighted spherical harmonics~\cite{Goldberg:1966uu}. $\Psi$ denotes a Hertz potential.

Here, since we are interested in the leading asymptotic behavior for large $r$,
the equation for $\psi_4$ is approximated as
\bea
r \psi_4 &=& \frac{1}{32} {\cal T}^4 r^5 {\bar \Psi} \,,
\eea
where ${\cal T} = (\partial_t - \partial_r^* )$.
In the above equation, the left hand side is written with respect to the retarded time
$t-r^*$. Therefore, using ${\cal T}^{-1} = (1/2) \int dt$, we have
\bea
\int \int \int \int dt \,dt \,dt \,dt \,(r \psi_4)
&=& \frac{1}{2} r^5 {\bar \Psi} \,.
\eea

For $\psi_0$, we ignore the term proportional to $M$
in order not to introduce the complex conjugation of ${\bar \Psi}^{\ell m}$,
and focus on the term $\edth^4 {\bar \Psi}$.
${\bar \Psi}$ is a spin-($-2$) function, i.e.,
\bea
\sum_{\ell m} \int \int \int \int dt \,dt \,dt \,dt \,(r \psi_4^{\ell m}) {}_{-2}Y_{\ell m}
&=& \frac{1}{2} r^5 \sum_{\ell m} {\bar \Psi}^{\ell m} {}_{-2}Y_{\ell m} \,.
\label{eq:psi4_app}
\eea
The operator $\edth$ on ${\bar \Psi}$ gives
\bea
\edth^4 {\bar \Psi} = \sum_{\ell m} (\ell-1)\ell(\ell+1)(\ell+2) {\bar \Psi}^{\ell m} {}_{2}Y_{\ell m} \,,
\eea
where there is no change in the $(t,\,r)$ dependence because $\edth$ acts only on the angular variables.
Although there may be a relation between $12 M \partial_t \Psi$ and 
the term proportional to $M$ in Eq.~(\ref{eq:rP4wP0_orig}) below
because both of the numerical factors are $3/2$,
we simply ignore it here
in order not to introduce the complex conjugation of ${\bar \Psi}^{\ell m}$.
This means that we consider an approximation,
\bea
\sum_{\ell m} \psi_0^{\ell m} {}_{2}Y_{\ell m} &=& \frac{1}{8}
\sum_{\ell m} (\ell-1)\ell(\ell+1)(\ell+2) {\bar \Psi}^{\ell m} {}_{2}Y_{\ell m} \,.
\label{eq:psi0_app}
\eea

Combining Eqs.~(\ref{eq:psi4_app}) and (\ref{eq:psi0_app}),
we have for each $(\ell,\,m)$ mode
\bea
\frac{1}{8}(\ell-1)\ell(\ell+1)(\ell+2)
\int \int \int \int dt \,dt \,dt \,dt \,(r \psi_4^{\ell m})
&=& \frac{1}{2} r^5 \psi_0^{\ell m} \,,
\eea
in the large $r$ limit and the above approximation.
Therefore, Eq.~(\ref{eq:asymtPsi4_2}) is rewritten as
\bea
\left. r\,\psi_4^{\ell m}\right|_{r=\infty} &=& r\,\psi_4^{\ell m}(t,r)
- \frac{(\ell -1)(\ell +2)}{2\,r} \int dt \,[r\,\psi_4^{\ell m}(t,r)]
\cr &&
+ \frac{(\ell^2+\ell-4)}{2\,\ell(\ell+1)\,r^2} \partial_t^2 [r^5\,\psi_0^{\ell m}(t,r)]
- \frac{3\,M}{2\,r^2} \int dt \,[r\,\psi_4^{\ell m}(t,r)]
+ {\cal O}(1/r^3) \,.
\label{eq:rP4wP0_orig}
\eea
Here, we have used $\psi_0^{\ell m}(t,r)$ extracted at a finite radius
because the error due to the use of finite extraction radii 
becomes higher order
in the large $r$ expansion. Since we have used an approximation
to derive Eq.~(\ref{eq:psi0_app}), for consistency, the $M$-dependent term 
should not be kept any more, i.e.,
\bea
\left. r\,\psi_4^{\ell m}\right|_{r=\infty} &=& r\,\psi_4^{\ell m}(t,r)
- \frac{(\ell -1)(\ell +2)}{2\,r} \int dt \,[r\,\psi_4^{\ell m}(t,r)]
\cr &&
+ \frac{(\ell^2+\ell-4)}{2\,\ell(\ell+1)\,r^2} \partial_t^2 [r^5\,\psi_0^{\ell m}(t,r)]
+ \textrm{[higher}\,\,\textrm{order]} \,.
\label{eq:asymtPsi4_2c}
\eea

This derivation is in an ideal situation where
we have assumed that there is no contribution from the other Weyl scalars,
the peeling theorem applies,
and we have used a low frequency $M \omega$ approximation.


\bibliographystyle{apsrev4-1}
\bibliography{../../../Bibtex/references}

\end{document}